\documentclass[amsmath,superscriptaddress,amssymb]{revtex4}

\usepackage{epsfig}

\begin{document}

\def\beqa{\begin{eqnarray}}
\def\eeqa{\end{eqnarray}}
\def\beqn{\begin{equation}}
\def\eeqn{\end{equation}}

\def\g{g}
\def\munu{{\mu\nu}}
\def\R{R}                 
\def\RE{E}                  
\def\PN{\Phi_N}
\def\PNa{{\Phi_a}}
\def\PP{{\Phi_P}}
\def\Pst{\phi}
\def\PNan{\delta \PN}
\def\h{h}                 
\def\Ps{\Phi^0}
\def\Pt{\Phi^1}
\def\Pa{\Phi^a}

\def\lr{{\bar \r}}        
\def\u{{u}}               %
\def\ub{{\bar \u}}        %
\def\gb{{\bar \g}}        %
\def\PbN{\delta {\bar \Phi}_N}   
\def\PbP{\delta {\bar \Phi}_P}
\def\REb{{\bar \RE}}

\def\k{k}                  
\def\ks{\mathbf{\k}}
\def\x{x}                  
\def\t{t}                  
\def\s{s}                  
\def\a{a}
\def\aan{\delta \a}
\def\r{r}                  
\def\rx{{\r_1}}
\def\ry{{\r_2}}
\def\ri{\rho}              
\def\th{\theta}            
\def\ang{\varphi}
\def\thn{\psi}
\def\thnan{\delta \thn}      

\def\T{T}

\def\pp{\pi}

\def\M{M}                  
\def\GN{G_N}
\def\GNa{G_a}
\def\GP{G_P}
\def\Gta{{\tilde G^a}}
\def\Gts{{\tilde G^0}}
\def\Gtt{{\tilde G^1}}
\def\GtN{{\tilde G_N}}
\def\GtNa{{\tilde G_a}}
\def\GtP{{\tilde G_P}}       
\def\cN{\zeta_N}
\def\cP{\zeta_P}

\def\c{c}                  
\def\vP{{v_P}}

\def\stand#1{\left[#1\right]_\mathrm{st}}
\def\unit#1{{\rm #1}}

\def\eprint#1{{\it Preprint} #1}

\title{Gravity tests and the Pioneer anomaly}
\author{ Marc-Thierry  Jaekel}

\affiliation{Laboratoire de Physique Th\'eorique de l'ENS,
24 rue Lhomond, F75231 Paris Cedex 05 \footnote{
Centre National de la Recherche Scientifique (CNRS), Ecole Normale Sup\'{e}rieure
 (ENS), Universit\'{e} Pierre et
Marie Curie (UPMC); email:jaekel@lpt.ens.fr}}

\author{Serge Reynaud }

\affiliation{Laboratoire Kastler Brossel, case 74, Campus Jussieu,
F75252 Paris Cedex 05 \footnote{CNRS, ENS, UPMC;
email:reynaud@spectro.jussieu.fr}}

\begin{abstract}
Experimental tests of gravity performed in the solar system 
show a good agreement with general relativity.
The latter is however challenged by the Pioneer anomaly which might be
pointing at some modification of gravity law at ranges 
of the order of the size of the solar system.
We introduce a metric extension of general relativity which, 
while preserving the equivalence principle, modifies 
the coupling between curvature and stress tensors
and, therefore, the metric solution in the solar system.
The ``post-Einsteinian extension'' replaces Newton gravitation constant 
by two running coupling constants, which depend on the scale and differ 
in the sectors of traceless and traced tensors, so that 
the metric solution is characterized by two gravitation potentials.
The extended theory has the capability to preserve compatibility with 
gravity tests while accounting for the Pioneer anomaly. 
It can also be tested by new experiments or, maybe, by having a new look
at data of already performed experiments.

PACS: 04.20.-q, 04.80.Cc \qquad\qquad LPTENS-05/35

\end{abstract}
\maketitle

\section{Introduction}

Most gravitation tests performed in the solar system show a good agreement 
with general relativity (GR). 
In particular, the equivalence principle, lying at the basis of GR, is one 
of the most accurately verified properties of nature \cite{Will01}.
This entails that the gravitational field has to be identified with 
the metric tensor $\g_\munu$ in a Riemannian space-time.
Then, the parametrized post-Newtonian (PPN) formalism allows one to give a 
quantitative form to the agreement of observations with the metric tensor predicted 
by GR, through its confrontation with a family of more general metric solutions.
Alternatively, GR can be tested by looking for hypothetical deviations 
of gravity force law from its standard form \cite{Fischbach98},
as predicted by unification models although not observed up to now. 

Besides these successes, GR is challenged by observations performed 
at galactic and cosmological scales. 
Anomalies have been known for some time to affect the rotation curves of galaxies.
They are commonly accounted for by keeping GR as the theory of gravity at galactic
scales but introducing unseen ``dark matter'' to reproduce the rotation curves 
\cite{Aguirre,Riess}. Anomalies have been seen more recently in the relation
between redshifts and luminosities for type II supernovae. They are usually 
interpreted as an unexpected acceleration of cosmic expansion due to the presence 
of some ``dark energy'' of completely unknown origin \cite{Perlmutter}. 
As long as the ``dark side'' of the universe is not observed through other means, 
these galactic and cosmic anomalies may also be interpreted as deviations 
from GR occuring at large scales \cite{Sanders02,Lue04,Turner04}.

The Pioneer anomaly constitutes a new piece of information in this puzzling context,
which may already reveal an anomalous behaviour of gravity at scales 
of the order of the size of the solar system \cite{Anderson98}.
The anomaly was discovered when Doppler tracking data from the Pioneer 10/11 probes 
were analyzed during their travel to the outer parts of the solar system.
After the probes had reached a quieter environment, after flying by Jupiter and
Saturn, a precise comparison of tracking data with predictions of GR confirmed
that the Doppler velocity was showing an anomaly varying linearly with elapsed time
(see Fig.~8 of \cite{Anderson02}).
The deviation may be represented as an anomalous acceleration directed towards 
the Sun with an approximately constant amplitude over a large range
of heliocentric distances (AU $\equiv$ astronomical unit)
\beqa
\label{Pioneer_acceleration}
a_{P} = (0.87 \pm 0.13) ~\unit{nm}~\unit{s}^{-2}\quad ,\quad 
20~\mathrm{AU}\lesssim\r\lesssim 70~\unit{AU}  &&
\eeqa

Though a number of mechanisms have been considered to this aim
\cite{Anderson02b,Anderson03,Nieto04,Turyshev04},
the anomaly has escaped up to now all attempts of explanation as a systematic
effect generated  by the spacecraft itself or its environment.
In particular, present knowledge of the outer part of the solar system does apparently 
preclude interpretations in terms of gravity \cite{Nieto05a}
or drag effects \cite{Nieto05} of ordinary matter. 
The inability of explaining the anomaly with conventional physics has given rise
to a growing number of new theoretical propositions. 
It has also motivated proposals for new missions designed to study the anomaly and 
try to understand its origin \cite{PAEM05}.
The importance of the Pioneer anomaly for space navigation already justifies it to be 
submitted to further scrutiny while its potential impact on fundamental physics, 
especially on gravitation theory, can no more be neglected.
The possibility that the Pioneer anomaly be the first hint of a modification of gravity 
law at large scales cannot be let aside investigations \cite{Bertolami04}. 
In this context, the compatibility of the Pioneer anomaly with other
gravity tests appears to be a key question. 

In order to discuss this point, we first recall that, though the interpretation 
of gravitation as the metric of space-time constitutes an extremely well tested
basis, the precise form of the coupling between space-time curvature and gravity 
sources can still be discussed \cite{Weinberg72}.  
Like the other fundamental interactions, gravitation may also be treated
within the framework of field theory \cite{Thirring,Feynman,Weinberg65}.
Radiative corrections due to its coupling to other fields then naturally lead 
to embed GR within the larger class of fourth order theories \cite{deWitt,Deser74,Capper74}.
Modifications are thus expected to appear \cite{Stelle,Sakharov,Adler}
and they may affect large length scales \cite{Nieto,Deffayet02,Dvali03,Gabadadze04}.
This suggests to consider GR as an effective theory of gravity valid at the length 
scales for which it has been accurately tested but not necessarily at smaller or
larger scales. 
Note that, in contrast to GR \cite{tHooft}, fourth order theories show renormalizability 
as well as asymptotic freedom at high energies \cite{Fradkin}.
Hence, they constitute a strong basis for extending the gravitation theory at scales 
not already constrained by experiments, for instance 
using renormalization group trajectories \cite{Reuter}. 
Renormalizability of these theories however comes with a counterpart, that is the 
problem of ghosts.
It has however been convincingly argued that this problem does not constitute a definitive 
deadend for an effective field theory valid in a limited scale domain \cite{Simon90}. 
In particular, the departure from unitarity is expected to be negligible at 
ordinary scales tested in present day universe \cite{Hawking02}.

In this paper, we will review the main features of a phenomenological framework
which has been recently developed \cite{JR05mpl,JR05cqg,JR05subm}. 
It relies upon a theory of gravitation lying in the vicinity of GR, the deviation
representing for example the radiative corrections due to the coupling of gravity
with other fields \cite{Jaekel95}. 
It is presented below in its linearized form, where its significance is more easily
given, and then in its full non linear version.
It is also interpreted as an extension of the PPN Ansatz with the Eddington 
parameters $\beta$ and $\gamma$ being functions of heliospheric distances rather
than mere constants.
The extended framework is shown to have the ability to account for Pioneer anomaly
while remaining compatible with other gravity tests. 
It also leads to the prediction of other anomalies related to Pioneer anomaly,
which can be tested by new experiments or, in some cases, by having a new look
at data of already performed experiments.

\section{Gravity tests in the solar system}
\label{tests}

GR provides us with an excellent theoretical description of gravitational 
phenomena in the solar system. In order to discuss the experimental evidences
in favor of this statement, we first recall a few basic features of this
description.

In order to apply the principle of relativity to accelerated motions, Einstein  
\cite{Einstein07,Einstein11} introduced what is now called the equivalence principle.
A weak form of this principle is expressed by the universality of free fall, 
a property reflecting the universal coupling of all bodies to gravitation. 
With Einstein, this property acquires a geometrical significance,  
gravitation fields being identified with the metric tensor $\g_{\mu\nu}$
while freely falling motions are  geodesics of the associated space-time. 
Universality of free fall is then a consequence of the metric nature of 
gravitation theory. 

The equivalence principle is one of the best ever tested properties of nature.
Potential violations are usually parametrized by a relative difference $\eta$ 
in the accelerations $a_1$ and $a_2$ undergone by two test bodies of different
compositions in free fall at the same location. Modern experiments constrain 
the parameter $\eta$ to stay below the $10^{-12}$ level.
These experiments test the principle at distances ranging from the
millimeter in laboratory experiments (\cite{Adelberger03} and references in)
to the sizes of Earth-Moon \cite{Williams96} or Sun-Mars orbit 
\cite{Hellings83,Anderson96}.

In order to obtain GR, it remains to write the equations determining the metric
tensor from the distribution of energy and momentum in space-time.
GR corresponds to a particular choice of the form of the coupling between curvature
tensor and stress tensor \cite{Einstein15,Hilbert,Einstein16}:
the Einstein curvature tensor $\RE_\munu$ is simply proportional to the stress tensor 
$\T_\munu$, the proportionality constant being related to the Newton gravitation 
constant $\GN$ inherited from classical physics
\beqa
\label{GR_gravitation_law}
\RE_\munu \equiv \R_\munu-{1\over2}\g_\munu R={8\pi\GN \over\c^4}\T_\munu
\eeqa
Note that this relation accounts in a simple manner for the fact that $\RE_\munu$
and $\T_\munu$ both have a null covariant divergence: the first property comes
with Riemannian geometry (Bianchi identities) while the second one expresses
conservation of energy and momentum and is a necessary and sufficient condition 
for motions of test masses to follow geodesics.

The metric tensor in the solar system is then deduced
by solving the Einstein-Hilbert equation (\ref{GR_gravitation_law}). 
Here we consider the simple case where the gravity source, \textit{i.e.} the Sun, 
is described as a point-like motion-less mass $\M$ so that
the metric is simply written in terms of the Newton potential $\Pst$.
The solution is conveniently written in terms of spherical coordinates
($\c$ denotes light velocity, $\t$ and $\r$ the time and radius,
$\th$ and $\ang$ the colatitude and azimuth angles)
with the gauge convention of isotropic spatial coordinates 
\beqa
\label{isotropic_metric}
d\s^2 &=& \g_{00} \c^2 d\t^2 + \g_{\r\r} \left( d\r^2  +
\r^2(d\th^2 + {\rm \sin}^2\th  d\ang^2) \right)\\
\label{GR_solar_metric}
\g_{00} &=& 1+2\Pst+2\Pst^2+\ldots\quad, \quad
\g_{rr} = - 1+2\Pst+\ldots \nonumber\\
\Pst &\equiv& -{\kappa\over\r}\quad, \quad 
\kappa \equiv {\GN\M\over\c^2}\quad, \quad 
\left\vert\Pst\right\vert \ll 1\nonumber
\eeqa

GR is usually tested through its confrontation with the family of 
parametrized post-Newtonian (PPN) metric tensors
introduced by Eddington \cite{Eddington} and then developed by  
several physicists \cite{Robertson,Schiff66,Nordtvedt68,WillNordtvedt72} 
\beqa
\label{PPN_0}
\g_{00} = 1 + 2 \alpha \Pst + 2 \beta \Pst^2 + \ldots \quad,\quad 
\g_{\r\r} = -1 + 2 \gamma \Pst + \ldots &&
\eeqa
The three parameters $\alpha$, $\beta$ and $\gamma$ are constants, the 
first of which can be set to unity by redefining Newton constant $\GN$.
Within the PPN family, GR corresponds to $\gamma=\beta=1$.
Anomalous values of $\gamma$ or $\beta$ differing from unity affect geodesic 
motions and can therefore be evaluated from a comparison of observations 
with predictions deduced from (\ref{PPN_0}). 

Experiments which have now been performed for more than four decades have led
to more and more strict bounds on the anomalies $\gamma-1$ and $\beta-1$. 
For example, Doppler ranging on Viking probes in the vicinity of Mars
\cite{Hellings83} and deflection measurements using VLBI astrometry \cite{Shapiro04} 
or radar ranging on the Cassini probe \cite{Bertotti03} give smaller and smaller
values of $\gamma-1$, with presently a bound of a few $10^{-5}$. 
Analysis of the precession of planet perihelions \cite{Talmadge88} and of 
the polarization by the Sun of the Moon orbit around the Earth 
\cite{LLR02} allow for the determination of linear superpositions 
of $\beta$ and $\gamma$, resulting now to $\beta-1$ smaller than a few $10^{-4}$.
These tests are compatible with GR with however a few exceptions, 
among which notably the anomalous observations recorded on Pioneer probes.
We will see below that this contradiction between Pioneer observations
and other gravity tests may be cured in an extended framework,
thanks to the fact that the anomaly $\gamma-1$ is no longer a constant but a 
function in this more general framework.

An alternative manner to test GR has been to check the $1/\r$ dependence of the Newton 
potential $\Pst$, that is also of the component $\g_{00}$ in (\ref{GR_solar_metric}).
Hypothetical modifications of its standard expression are usually parametrized 
in terms of an additional  Yukawa potential depending on two parameters, 
the range $\lambda$ 
and the amplitude $\alpha $ measured with respect to Newton potential $\Pst$
\beqa
\label{Yukawa_perturbation}
\PN (\r) =\Pst(\r) \left( 1+\alpha e^{-\frac\r\lambda}\right)  && 
\eeqa
The presence of such a Yukawa correction has been looked for at various distances
ranging from the millimeter in laboratory experiments \cite{Adelberger03} 
to the size of planetary orbits \cite{Coy03}.
The accuracy of short range tests has been recently improved, as gravity experiments 
were pushed to smaller distances \cite{Hoyle,Long,Chiaverini} and as Casimir forces, 
which become dominant at submillimeter range, were more satisfactorily taken into 
account \cite{Bordag01,Lambrecht02,Decca03,Chen04}.
On the other side of the distance range, long range tests of the Newton law are 
performed by monitoring the motions of planets or probes in the solar system. 
They also show an agreement with GR with a good accuracy for ranges of the order 
of the Earth-Moon \cite{Williams96} or Sun-Mars distances 
\cite{Reasenberg79,Hellings83,Anderson96,Kolosnitsyn03}. 
When the whole set of results is reported on a global figure (see fig.1 in
\cite{JR04} reproduced thanks to a courtesy of the authors of \cite{Coy03}),
it appears that windows however remain open for violations of the standard
form of Newton force law at short ranges, below the millimeter, as well as
long ones, of the order of or larger than the size of the solar system.

One merit of the latter tests is to shed light on a potential
scale dependence of violations of GR. As a specific experiment is only sensitive
to a given range of distances, this has to be accounted for,
especially in the context, recalled in the Introduction, where doubts arise
about the validity of GR at galactic or cosmic scales.
In order to discuss this scale dependence, it is worth rewriting the Yukawa
perturbation (\ref{Yukawa_perturbation}) in terms of a running constant
replacing Newton gravitation constant.
To this aim, we introduce the expression of the potential $\PN[\mathbf{k}]$ 
in Fourier space, with $\mathbf{k}$ the spatial wavevector, and relate it 
to a coupling constant $\widetilde{G}_N[\mathbf{k}]$ 
\beqa
-\mathbf{k}^{2}\PN \left[ \mathbf{k}\right] \equiv 4\pi \frac{\widetilde{G}_N%
\left[ \mathbf{k}\right] M}{c^{2}}\qquad ,\qquad \widetilde{G}_N\left[ 
\mathbf{k}\right] =\GN\left( 1+\alpha\frac{\mathbf{k}^{2}}
{\mathbf{k}^{2}+\frac{1}{\lambda ^{2}}}\right)   &&
\label{Running}
\eeqa
The left-hand equation extends the standard Poisson equation 
($-\mathbf{k}^{2}$ is the Laplacian operator), with the constant
$\GN$ replaced by a ``running coupling constant'' $\widetilde{G}_N[\mathbf{k}]$ 
which depends on spatial wavevector $\mathbf{k}$.

Note that the Yukawa correction (\ref{Yukawa_perturbation}) gives rise in the domain 
$\r\ll\lambda$ to a perturbation $\delta\PN$ which is linear in the distance 
$\sim-(\alpha/2\lambda^2)(\kappa\r)$.
In equation (\ref{Running}) equivalently, the correction of the running constant
scales as $\delta \widetilde{G}_N \sim (\alpha/\lambda^2)(\GN/\mathbf{k}^2)$
in the domain $\vert\mathbf{k}\lambda\vert\gg 1$.
In fact, experimental constraints obtained at scales of the order of the size of the
solar system can be written as bounds on the combination $\alpha/\lambda ^2$, so that 
they can be translated into bounds on the anomalous acceleration $\partial_\r\delta\PN$
or, equivalently, on $\mathbf{k}^2\delta\widetilde{G}_N$. 
It is worth emphasizing that these bounds result in allowed anomalous accelerations
which remain 2000 times too small
to account for the Pioneer anomaly \cite{JR04}. 
In other words, the Pioneer anomaly cannot be due to a modification of Newton law, as
such a modification would be much too large to remain unnoticed
by planetary tests. 
Anew, this contradiction between Pioneer observations and other gravity tests may be 
cured by the extended framework studied below, now thanks to the fact that the existence
of an anomalous space-dependent potential will not only be considered for the metric 
component $\g_{00}$ but also for $\g_{\r\r}$.

To summarize this section, 
tests performed on gravity in the solar system confirm its metric character
and provide strong evidence in favor of gravitation theory being very close to
GR. They however still leave room for alternative metric theories, 
which deviate from GR in a specific way. 
Anomalies in the metric components
remain allowed, as long as  they modify  spatial dependencies
without strongly affecting the time component $\g_{00}$.
It is shown in next sections that such extensions of GR 
may in fact arise naturally, in particular when 
effects of radiative corrections are taken into accont. 

\section{Linearized gravitation theory}

We come now to the description of the ``post-Einsteinian'' extension of GR.
We first repeat that tests performed at various length scales have showed
that the equivalence principle (EP) was preserved at a higher accuracy level, 
10$^{-12}$, than the EP violation which would be needed to account for the
Pioneer anomaly. As a matter of fact, the standard Newton acceleration 
at 70~UA is of the order of 1~$\mu$m~s$^{-2}$ while the Pioneer anomaly is 
of the order of 1~nm~s$^{-2}$, which would correspond to a violation level of 
the order of 10$^{-3}$.
This does not mean that EP violations are excluded, and they are indeed 
predicted by unification models \cite{Damour,Overduin00}.
However, any such violations are bound to occur at a lower level than needed
to affect the Pioneer anomaly. Hence, EP violations will be ignored in the following
and we shall restrict our discussion to a
confrontation of GR with alternative metric theories.

Furthermore, neither PPN extensions of GR nor mere modifications of Newton force laws
have the ability to account for the Pioneer anomaly (see the previous section).
However, such extensions do not cover the totality of possible extensions of GR.
In particular, there exist extended metric theories characterized by the existence of
two gravitation potentials instead of a single one \cite{JR05mpl,JR05cqg,JR05subm}.
The first one merely represents a modified Newton potential while the second one
can be understood in terms of a space-dependent PPN parameter $\gamma$.
In this larger family of extensions, there is enough room available for accomodating 
the Pioneer anomaly while preserving compatibility with other gravity tests.
Let us stress that this larger family is not introduced as an {\it adhoc} solution
to the Pioneer anomaly. It emerges  as the natural extension of GR induced by 
radiative corrections due to the coupling of gravity with other fields,
and some phenomenological consequences were explored \cite{Jaekel95}
before noticing that they included  Pioneer-like anomalies 
\cite{JR05mpl,JR05cqg,JR05subm}. 
In order to present these ideas in a simple manner, we will start with the linearized
version of gravitation theory, which is approximately valid for describing Pioneer-like 
probes having escape motions in the outer solar system \cite{JR05mpl,JR05cqg}. 
We will then present some salient features of the non linear theory \cite{JR05subm}.

In the linearized treatment, the metric field may be represented as a small perturbation 
$\h_\munu$ of Minkowski metric $\eta_\munu$ 
\beqa
&&\g_\munu = \eta_\munu + \h_\munu \\
&&\eta_\munu = {\rm diag}(1, -1, -1, -1) 
\quad,\quad \left\vert \h_\munu \right\vert \ll 1 \nonumber
\eeqa
The field $\h_\munu$ is a function of position $x$ in spacetime or, 
equivalently in Fourier space, of wavevector $k$
\beqa
\h_{\mu\nu}(\x) \equiv \int {d^4\k \over (2\pi)^4}e^{-i\k\x} \h_{\mu\nu}[\k]
\eeqa 
Gauge invariant observables of the metric theory, {\it i.e.} quantities which do not depend 
on a choice of coordinates, are given by curvature tensors. In the linearized theory, 
{\it i.e.} at first order in $\h_{\mu\nu}$, Riemann, Ricci, scalar and Einstein curvatures are 
written in momentum representation as 
\beqa
\label{curvatures}
&&\R_{\lambda\mu\nu\rho} = \frac{
\k_\lambda\k_\nu \h_{\mu\rho} - \k_\lambda\k_\rho \h_{\mu\nu}
- \k_\mu\k_\nu \h_{\lambda\rho} + \k_\mu\k_\rho \h_{\lambda\nu}}2\\
&&\R_{\mu\nu} = {\R^\lambda}_{\mu\lambda\nu}\quad, \quad \R = {\R^\mu}_\mu\quad,\quad
\RE_{\mu\nu} = \R_{\mu\nu} -\eta_{\mu\nu}{\R\over2}\nonumber
\eeqa
We use the sign conventions of \cite{Landau}, indices being raised or 
lowered using Minkowski metric. 

These curvature fields are similar to the gauge invariant electromagnetic fields
of electrodynamics so that, while being supported by its geometrical interpretation, 
GR shows essential similarities with other field theories \cite{Feynman,deWitt}.
This suggests that GR may be considered as the low energy effective limit
of a more complete unified theory \cite{Sakharov,Adler} which should describe
the coupling of gravity with other fields.
In any case, this theory should contain radiative corrections to the graviton propagator,
leading to a modification of gravitation equations (\ref{GR_gravitation_law}) and
to a momentum dependence of the coupling between curvature and stress tensors.
In the weak field approximation, it is easily seen that Einstein tensor, which is
divergenceless, has a natural decomposition on the two sectors corresponding to 
different conformal weights \cite{Jaekel95}, that is also on traceless (conformal weight 0)
and traced components (conformal weight 1). 
The general coupling between curvature and stress tensors can be written in terms of a 
linear response function constrained by the transversality condition 
\beqa
\label{general_gravitation_law}
\RE_\munu [k] = \chi_{\mu\nu\lambda\rho} [k] ~\T^{\lambda\rho} [k] 
\quad,\quad k^\mu \chi_{\mu\nu\lambda\rho} [k] = 0
\eeqa

We consider as above the isotropic and stationary situation with a point-like and 
motion-less Sun of mass $M$. We then deduce that the general coupling 
(\ref{general_gravitation_law}) is described by two running constants $\Gts$ and $\Gtt$,
depending on the spatial wavevector $\mathbf{k}$ and living 
in the two sectors (0) and (1), 
so that gravitation equations (\ref{general_gravitation_law}) become \cite{JR05mpl,JR05cqg}
\beqa
\label{isotropic_Einstein}
&&\RE_\munu [k] \equiv 2\pi \delta(k_0) \RE_\munu [\mathbf{k}], \qquad   \pi _{\mu \nu }
\equiv\eta _{\mu \nu }-\frac{k_{\mu }k_{\nu }}{k^{2}}\nonumber\\
&&E_\munu [\mathbf{k}] = \pi _\mu^0 \pi_\nu^0\, \Gts [\mathbf{k}] 
\frac{8\pi M}{c^2} +  \pi _\munu\pi ^{00}\, 
\frac{\Gtt [\mathbf{k}] - \Gts [\mathbf{k}] }{3}\,
\frac{8\pi M}{c^2} 
\eeqa
The Newton gravitation constant $\GN$ in (\ref{GR_gravitation_law})
has been replaced in (\ref{isotropic_Einstein}) by two running 
coupling constants $\Gts$ and  $\Gtt$ which are related through Poisson-like 
equations to two potentials $\Ps$ and $\Pt$ (compare with (\ref{Running}))
\beqa
\label{inv_grav_equations}
&&-\ks^2 \Pa[\ks] =  \Gta[\ks]{4\pi \M \over \c^2} 
\quad,\quad a=0,1 
\eeqa
These two potentials determine the metric, that is the solution of the modified equations 
(\ref{isotropic_Einstein}), written here with spatial isotropic coordinates
\beqa
\label{polar_metric_NP}
\g_{00} &=& 1 + 2 \PN \quad,\quad \PN \equiv {4 \Ps - \Pt \over 3} \\
\g_{\r\r} &=& -(1 - 2 \PN + 2\PP) \quad,\quad 
\PP \equiv {2 (\Ps - \Pt) \over3} \nonumber
\eeqa
$\PN$ is defined from the difference $(\g_{00}-1)$ and identified as an extended Newton potential  
while $\PP$ is defined from $(-\g_{00} \g_{\r\r}-1)$ and interpreted as measuring the difference 
between the potentials $\Ps$ and $\Pt$ in the two sectors of traceless and traced curvatures.
As $\Ps$ and $\Pt$, $\PN$ and $\PP$ obey Poisson equations
with running constants $\GtN$ and $\GtP$ written as
linear combinations of $\Gts$ and $\Gtt$ 
\beqa
&&-\ks^2 \PNa[\ks] =  \GtNa[\ks]{4\pi \M \over \c^2} 
\quad,\quad a= N,P  \\
&&\GtN \equiv {4 \Gts - \Gtt \over 3} \quad,\quad 
\GtP \equiv {2 (\Gts - \Gtt) \over3}  \nonumber
\eeqa

Standard Einstein equation is recovered when the running constants $\Gts$ and  $\Gtt$
are momentum independent and equal to each other, that is also when 
\beqa
&&\stand{\GtN} \equiv  \GN \quad,\quad \stand{\GtP} = 0 \nonumber\\
&&\stand{\PN(\r)} \equiv  \Pst(\r) \quad,\quad 
\stand{\PP(\r)} = 0 
\eeqa
The two potentials $\PNa$ will be written as sums of these standard expressions and
anomalies which, according to the discussions of the previous section,
will remain small
\beqa
\label{two_potentials}
\PNa (\r) \equiv \stand{\PNa(\r)} + \delta\PNa (\r) \quad,\quad 
\left| \delta\PNa (\r) \right| \ll 1 &&
\eeqa

\section{Non linear gravitation theory}

Before embarking in the discussion of phenomenological consequences of these
anomalous potentials, let us recall briefly that the linearized theory
presented in the preceding section can be transformed into a full non linear
theory. The linearized theory will indeed be sufficient to discuss the 
anomalous acceleration of Pioneer probes as well as potential effects
on light-like waves \cite{JR05mpl,JR05cqg} but the non linear theory
will be needed to address the case of planetary tests \cite{JR05subm}.
The relation between metric and curvature tensors now takes a non linear form
going beyond (\ref{curvatures}) which was valid only at first order around
Minkowski space-time. We may nevertheless simplify this relation by working 
at first order in deviations from standard Einstein theory \cite{JR05subm}.

To this aim, we write the metric, now in terms of 
Schwartzschild coordinates \cite{MTW} 
\beqa
\label{Schwartzschild_isotropic_metric}
&&d\s^2 = \gb_{00}(\lr)\c^2 d\t^2 + \gb_{\r\r}(\lr)d\lr^2 
- \lr^2 \left(d\theta^2 + {\rm{sin}}^2\theta d\varphi^2\right)\nonumber\\
&&\gb_\munu(\r) \equiv \stand{\gb_\munu(\r)} + \delta\gb_\munu (\r)  \quad,\quad 
\left| \delta\gb_\munu(\r) \right| \ll 1 
\eeqa
The standard GR solution is then treated exactly 
\beqa
\label{metric_st}
&&\stand{\gb_{00}} = 1-2\kappa\ub =  -{1\over\stand{\gb_{\r\r}}} \quad , \quad 
\ub\equiv{1\over\lr} 
\eeqa
while the anomalous metric components are taken into account at first order. 
Proceeding in this manner, it is possible to define in the non linear theory
two potentials $\PbN$ and $\PbP$ which generalize (\ref{two_potentials})
\beqa
\label{Schwartzschild_solution}
&&\delta\gb_{\r\r} = {2\ub\over(1-2\kappa\ub)^2}(\PbN-\PbP)^{\prime} \quad, \quad
f^\prime \equiv \partial_\ub f \nonumber\\
&&\delta\gb_{00} = 2(1-2\kappa\ub)
\int{\PbN^\prime -2\kappa\ub\PbP^\prime\over(1-2\kappa\ub)^2}d\ub 
\eeqa
In the linearized approximation, corrections in $\kappa\ub$ are disregarded
and the simple relations of the preceding section are recovered.
In the general case, equations (\ref{Schwartzschild_solution}) fully describe 
non linear effects of the Newton potential $\kappa\ub$. 
The precise form of the non linear version (\ref{Schwartzschild_solution}) has
been chosen so that potentials are related in a simple way to the corresponding anomalous
Einstein curvatures 
\beqa
\label{curvature_solution}
\delta\REb^0_0 &\equiv& 2\ub^4(\PbN -\PbP)^{\prime\prime}\nonumber\\ 
\delta\REb^\r_\r &\equiv& 2\ub^3\PbP^\prime
\eeqa

At this stage, it is worth noticing that the PPN metric \cite{Will01} may be recovered 
as a particular case of the more general extension (\ref{Schwartzschild_solution}).
This particular case corresponds to the following expressions of anomalous potentials 
and anomalous Einstein curvatures \cite{JR05subm}
\beqa
\label{PPN_potentials}
&&\PbN = (\beta-1)\kappa^2\ub^2 + O(\kappa^3\ub^3 ) \quad,\quad \mathrm{[PPN]}\nonumber\\
&&\PbP = (\gamma-1)\kappa\ub +O(\kappa^2\ub^2) 
\eeqa
\beqa
\label{PPN_curvatures}
&&\delta\RE^0_0 = \ub^2 O( \kappa ^2 \ub^2) \quad,\quad \mathrm{[PPN]}\nonumber\\
&&\delta\RE^\r_\r = \ub^2 \left(2 (\gamma-1) \kappa  \ub +O( \kappa ^2 \ub^2) \right) 
\eeqa
Note that the PPN metric already shows an anomalous behaviour of Einstein curvatures 
which have non null values apart from the gravity source. 
This is the case for $\delta\RE^\r_\r$ at first order in $\kappa$, and for $\delta\RE^0_0$ 
at higher orders.
Relations (\ref{curvature_solution}) thus extend this anomalous behaviour to more 
general dependences of the curvatures $\delta\RE^0_0$ and $\delta\RE^\r_\r$. 
Similar statements apply as well for anomalous potentials $\PbN$ and $\PbP$,
 which generalize the specific dependence of PPN potentials (\ref{PPN_potentials})
 where $\beta-1$ and $\gamma-1$
are constants. 
In other words, the post-Einsteinian metric (\ref{Schwartzschild_solution}) can be 
thought of as an extension of PPN metric where $\beta-1$ and $\gamma-1$
are no longer constants but rather functions of space.

\section{Phenomenological consequences}

As already discussed, the new phenomenological framework is characterized by two
anomalous potentials: the first one $\PbN$ is a modification of Newton potential  
while the second one $\PbP$ represents the difference of gravitational couplings
in the two sectors of traceless and traced curvatures.
The first potential is not able by itself to explain the Pioneer anomaly: its
anomalous part is indeed bound by planetary tests to be much smaller than would
be needed to account for the Pioneer anomaly \cite{JR04}.
This is why we will focus the attention in the following on the second potential
which can produce a Pioneer-like anomaly for probes on escape trajectories in
the outer solar system \cite{JR05mpl,JR05cqg}.
This second potential can also be understood as promoting the PPN parameter
$\gamma$ to the status of a space dependent function and it has therefore
other consequences which have to be evaluated with great care:
It is clear that the modification of GR needed to produce the 
Pioneer anomaly should not spoil its agreement with other gravity tests.

We first discuss the effect of the second potential on Doppler tracking of
Pioneer-like probes. To this aim, we calculate the Doppler velocity
taking into account the perturbations on probe motions as well as on
light propagation between stations on Earth and probes.
We then write the time derivative of this velocity as an acceleration $a$
and finally subtract the expression obtained in standard theory from
that obtained in extended one. We thus obtain the prediction of the
post-Einsteinian extension \cite{JR05mpl,JR05cqg} for the
Pioneer anomalous acceleration $\aan \equiv a - \stand{a}$.
In a configuration similar to that of Pioneer 10/11 probes, 
which follow nearly radial trajectories with a kinetic energy much larger
than their potential energy, this anomalous acceleration takes the simplified form
\beqa
\label{Pioneers_acceleration}
\aan &\simeq& 2 {d \delta\PP \over d\r} \vP^2 
\eeqa
Thus, an anomaly in Doppler tracking of Pioneer-like probes is a direct 
consequence of the presence of the second potential $\delta\PP$. Note that the
anomalous acceleration comes out as proportional to the kinetic energy, 
which is a remarkable prediction of the new framework.
Data on probes with very different kinetic energies (unfortunately not available at 
the moment) could thus be used to confirm or infirm this prediction.

Using the known velocity of the Pioneer probes ($\vP \sim 12 {\rm{km~s}}^{-1}$),
and identifying the acceleration (\ref{Pioneers_acceleration}) with the recorded 
Pioneer anomaly (\ref{Pioneer_acceleration}),
we deduce the value of the derivative $d\delta\PP / d\r$ in the outer solar system.
The constancy of recorded anomaly over a large range of distances  agrees with 
a simple parametrization of the second potential \cite{JR05mpl}
\beqa
\label{Pioneer_simple_model}
&&\delta\PP(\r) \equiv -{\GP \M \over \r\c^2} + {\cP\M\r\over\c^2} \nonumber\\
&&\left\vert{\GP\over\GN}\right\vert\ll 1\quad,\quad 
\cP \M \sim 0.25~ \rm{m~s}^{-2} 
\eeqa
This value of the parameter $\cP\M$ is much larger than that allowed
for the parameter $\cN\M$ which could be defined on the first potential \cite{JR05mpl}.
This shows in a clear manner how the second potential $\delta\PP$ opens the
possibility to account for the Pioneer anomaly. 
It has  to be kept in mind for the forthcoming discussions that the simple model 
(\ref{Pioneer_simple_model}) does not need to be exact in the whole solar system. 
However, the expression of $\delta\PP$ can generally be given the form 
 (\ref{Pioneer_simple_model}),
provided $\cP$ denotes a function of the heliocentric distance.

We come now to the discussion of the effects of the second potential $\delta\PP$ on the
propagation of light rays, which can be done in the linearized theory.
Considering in particular deflection experiments usually devoted to the 
determination of Eddington parameter $\gamma$, we obtain the following expression
for the anomaly $\thnan$ (with respect to GR) of the deflection angle  
of rays grazing the surface of the Sun \cite{JR05cqg} (the same decomposition 
as in (\ref{Pioneer_simple_model}) is used for $\delta\PP$) 
\beqa
\label{anomalous_time_delay}
\label{anomalous_angle}
&&\thnan \simeq  -\kappa {\partial \over \partial \ri}  
\left( \delta\gamma(\ri)~{\rm{ln}}{4 \rx\ry\over \ri^2} \right)
\nonumber\\
&&\delta \gamma(\ri) 
= -{\GP\over\GN} +{\cP(\ri)\ri^2\over 2\GN}
\eeqa
Terms which are not amplified near occultation have been neglected;
$\rx$ and $\ry$ correspond to the heliocentric distances of the receiver and emitter, 
and $\ri$ is the distance of closest approach of the light ray to the Sun; 
$\delta \gamma(\ri)$ is a range dependent anomalous part in Eddington parameter $\gamma$.
These expressions are reduced to PPN ones when the function $\cP$ vanishes.
Otherwise, they show that Eddington deflection tests could reveal the presence 
of $\delta\PP$ through a space dependence of the parameter $\gamma$.

We conclude this survey of phenomenologocal consequences of the new framework
by discussing planetary tests and, in particular, those involving the perihelion precession
of planets. As the latter are known to depend on the two PPN parameters $\beta$ and
$\gamma$, this discussion has to be presented in the context of the non linear theory. 
Still focusing the attention on the effects of the second
potential, one thus obtains the following expression for the anomaly $\delta\Delta \varphi$
 (with respect to GR)
of the perihelion precession \cite{JR05subm}
\beqa
{\delta\Delta \varphi\over 2\pi} \simeq \ub\left(\ub\PbP\right)^{\prime\prime} 
+ {e^2 \ub^2\over8}\left(\ub^2\PbP^{\prime\prime}+\ub\PbP^\prime\right)
^{\prime\prime} &&
\eeqa
This expression has been truncated after leading and sub-leading orders in the eccentricity 
$e$; the function $\PbP$ and its derivatives have thus to be evaluated at the inverse radius
$\ub$ of the nearly circular ($e\ll1$) planetary orbit.
As the leading order vanishes for a contribution $\cP\M\r/\c^2$ with $\cP$ constant,
the main result is thus proportional to $e^2$ in this case.
This means that perihelion precession of planets could be used as a sensitive probe
of the value and variation of $\cP$ for distances corresponding to the radii of
planetary orbits \cite{JR05subm}.

\section{Discussion}

Gravity tests which have been performed up to now in the solar system firmly support 
the validity of the equivalence principle, that is also the metric nature of gravitation.
They also strongly indicate that the actual gravitation theory should be very close to GR.
Nonetheless, these tests still leave room for alternative metric theories of gravitation, 
and the anomaly observed on the trajectories of the Pioneer 10/11 probes may well be
a first indication of a modification of gravity law in the outer part of the solar system. 
This possibility would have such a large impact on fundamental physics, astrophysics
and maybe cosmology that it certainly deserves further investigations.
We have discussed in the present paper a new extension of GR which allows one to address
these questions in a well defined theoretical framework \cite{JR05cqg,JR05subm}.

When its radiative corrections are taken into account, GR appears as imbedded 
in a family of metric theories characterized, at the linearized level, 
by two running coupling constants which replace the single Newton gravitation constant 
or, equivalently, by two potentials which replace the standard Newton potential.
When applied to the solar system, this post-Einsteinian extension of GR
leads to a phenomenological framework which has the ability to make the Pioneer
anomaly compatible with other gravity tests. 
Precisely, the first potential $\PN$ remains close to its standard Newtonian form in 
order to fit planetary tests but the second potential $\PP$ opens a phenomenological
freedom which can be understood as an Eddington parameter $\gamma$ differing from
unity, as in PPN metric, with now a possible space dependence.

In order to confirm, or infirm, the pertinence of this framework with respect
to gravity tests in the solar system, it is now necessary to re-analyze the motions
of massive or massless probes in this new context.
Contrarily to what has been done here, it is particularly important to take into
account the effects of anomalous potentials $\delta\PN$ and $\delta\PP$ simultaneously.
Let us scan in the last paragraphs of this paper some ideas which look
particularly promising.

The main novelty induced by the second potential $\delta\PP$ is to produce an anomaly
on Doppler tracking of Pioneer-like probes having highly eccentric motions
in the outer solar system. As already discussed, if the recorded anomaly is
identified with this effect, one deduces the value of the derivative $d\delta\PP/d\r$
of the second potential at the large distances explored by Pioneer probes.
A natural idea is therefore to confront the more detailed prediction deduced
from the new theory \cite{JR05cqg} against the larger set of data which will soon 
be available \cite{Nieto05early,TuryshevO5grex}.
It is particularly clear that the eccentricity of the orbits plays a key role in 
the evaluation of the Pioneer anomaly: it takes large values for Pioneer-like motions
which sense $\delta\PP$ whereas it is zero for circular orbits which do not.
This suggests to devote a dedicated analysis to the intermediate situation,
not only for the two categories of bound and unbound orbits, but also for the 
flybies used to bring Pioneer-like probes from the former category to 
the latter one.
While Pioneer probes may sense the second potential at the large heliocentric
distances they are exploring, planets or planetary probes may feel its presence 
at distances of the order of the astronomical unit.
Then, it would be worth studying planetary probes on elliptical
orbits, for example on transfer orbits from Earth to Mars or Jupiter.
Another natural target for such a study is LISA with its three crafts
on elliptical orbits \cite{Lisa}. 

The second potential $\delta\PP$ also affects the  propagation of light waves, and it could 
thus be detected as a range dependence of the anomalous Eddington parameter $(\gamma-1)$
to be seen for example in deflection experiments.
This might already be attainable through a reanalysis of existing data,
given by the Cassini experiment \cite{Bertotti03},
VLBI measurements \cite{Shapiro04} or HIPPARCOS \cite{Hipparcos}.
It may also be reached in the future by higher accuracy Eddington tests,
as for example the LATOR project \cite{LATOR05}, or global mapping of deflection
over the sky (GAIA project \cite{Gaia}). 
Reconstruction of the dependence of $\gamma$ versus the impact
parameter $\ri$ would directly provide the space dependence of the second
potential $\delta\PP$.
This would then either produce a clear signature of
the new framework presented in this paper or put constraints on the presence
of the second potential at small heliocentric distances.


\begin{references}
\bibitem{Will01} Will C.M., \textit{Theory and experiment in gravitational physics} 
(Cambridge U. P., Cambridge, 1993); \textit{Living Rev. Rel.} \textbf{4} (2001) 4.
\bibitem{Fischbach98}  Fischbach E. and Talmadge C., 
\textit{The Search for Non Newtonian Gravity} (Springer Verlag, Berlin, 1998). 
\bibitem{Aguirre}  Aguirre A., Burgess C.P., Friedland A. and Nolte D.,
{\it Class. Quantum Grav.} \textbf{18} (2001) R223.
\bibitem{Riess}  Riess A. G., Filippenko A. V., Challis P. \etal,
{\it Astron. J.} \textbf{116} (1998) 1009.
\bibitem{Perlmutter}  Perlmutter S., Aldering G., Goldhaber G. \etal,
{\it Astrophys. J.} \textbf{517} (1999) 565;
Perlmutter S., Turner M. S. and White M.,
\textit{Phys. Rev. Lett.} \textbf{83} (1999) 670.
\bibitem{Sanders02} Sanders R. H. and McGaugh S. S., 
{\it Annu. Rev. Astron. Astrophys.} \textbf{40} (2002) 263.
\bibitem{Lue04}  Lue A., Scoccimari R. and Starkman G., 
{\it Phys. Rev.} {\bf D 69} (2004) 044005.
\bibitem{Turner04}  Carroll S.M., Duvvuri V., Trodden M. and Turner M.S., 
{\it Phys. Rev.} {\bf D 70} (2004) 043528.
\bibitem{Anderson98} Anderson J.D., Laing P.A., Lau E.L. \etal, 
{\it Phys. Rev. Lett.} {\bf 81} (1998) 2858.
\bibitem{Anderson02} Anderson J.D., Laing P.A., Lau E.L. \etal, 
{\it Phys. Rev.} {\bf D 65} (2002) 082004.
\bibitem{Anderson02b} Anderson J.D., Nieto M.M. and Turyshev S.G., \textit{%
Int. J. Mod. Phys.} \textbf{D11}, 1545 (2002).
\bibitem{Anderson03} Anderson J.D., Lau E.L., Turyshev S.G. \etal, 
\textit{Mod. Phys. Lett.} \textbf{A17} (2003) 875.
\bibitem{Nieto04} Nieto M.M. and Turyshev S.G., \textit{Class. Quantum Grav.} 
\textbf{21} (2004) 4005.
\bibitem{Turyshev04} Turyshev S.G., Nieto M.M. and Anderson J.D.,
{\it 35th COSPAR Scientific Assembly} (2004) \eprint{gr-qc/0409117}.
\bibitem{Nieto05a} Nieto M. M., \textit{ Phys. Rev.} \textbf{D 72} (2005) 083004.
\bibitem{Nieto05} Nieto M. M., Turyshev S. G. and Anderson J. D., \textit{Phys.
Lett.B} \textbf{613} (2005) 11.
\bibitem{PAEM05} The Pioneer Explorer Collaboration: H. Dittus, S.G. Turyshev \etal
{\it A Mission to Explore the Pioneer Anomaly} \eprint{gr-qc/0506139}.
\bibitem{Bertolami04}
Bertolami O. and Paramos J., \textit{Class. Quantum Grav.} \textbf{21} (2004) 3309;
see also  \eprint{astro-ph/0408216} and  \eprint{gr-qc/0411020}.
\bibitem{Weinberg72} Weinberg S., {\it Gravitation and Cosmology} (John Wyley and 
Sons, New York, 1972).
\bibitem{Thirring} Thirring W. E., {\it Ann. of Phys.} {\bf 16}  (1961) 96.
\bibitem{Feynman} Feynman R.P., {\it Acta Phys. Polonica} {\bf 24} 711 (1963).
\bibitem{Weinberg65} Weinberg S., {\it Phys. Rev.} {\bf 138} B988 (1965).
\bibitem{deWitt} Utiyama R. and De Witt B., {\it J. Math. Phys.} 
{\bf 3} (1962) 608. 
\bibitem{Deser74}  Deser S. and van Nieuwenhuizen P., 
{\it Phys. Rev.} {\bf D 10} (1974) 401.
\bibitem{Capper74}  Capper D.M., Duff M.J. and Halpern L., 
{\it Phys. Rev.} {\bf D 10} (1974) 461.
\bibitem{Stelle}  Stelle K.S., {\it Phys. Rev.} {\bf D 16} (1977) 953;
{\it Gen. Rel. Grav.} {\bf 9} (1978) 353.
\bibitem{Sakharov} Sakharov A. D., {\it Doklady Akad. Nauk SSSR} 
{\bf 177} (1967) 70 [{\it Sov. Phys. Doklady} {\bf 12} 1040]. 
\bibitem{Adler}  Adler R.J., {\it Rev. Modern Phys.} {\bf 54} (1982) 729.
\bibitem{Nieto} Goldman T., P\'erez-Mercader J., Cooper F. and Nieto M. M.,
{\it Phys. Lett.} {\bf B 281} 219.
\bibitem{Deffayet02} Deffayet C., Dvali G., Gabadadze G. and Vainshtein A., 
{\it Phys. Rev.} {\bf D 65} (2002) 044026.
\bibitem{Dvali03} Dvali G., Gruzinov A., Zaldarriaga M., 
{\it Phys. Rev.} {\bf D 68} (2003) 024012.
\bibitem{Gabadadze04} Gabadadze G. and Shifman M., 
{\it Phys. Rev.} {\bf D 69} (2004) 124032.
\bibitem{tHooft} t'Hooft G. and Veltman M., {\it Ann. Inst. H. Poincar\'e} 
{\bf A 20} (1974) 69.
\bibitem{Fradkin} Fradkin E. S. and Tseytlin A. A., {\it Nucl. Phys.} {\bf B 201}
(1982) 469.
\bibitem{Reuter} Lauscher O. and Reuter M., \textit{Class. Quantum Grav.} \textbf{19}
(2002) 483.
\bibitem{Simon90} Simon J. Z., {\it Phys. Rev.} {\bf D 41}
(1990) 3720.
\bibitem{Hawking02} Hawking S. W. and Hertog T., {\it Phys. Rev.} {\bf D 65}
(2002) 103515.
\bibitem{JR05mpl}   Jaekel M.-T. and Reynaud S.,
{\it Mod. Phys. Lett.} \textbf{A20} (2005) 1047.
\bibitem{JR05cqg}   Jaekel M.-T. and Reynaud S.,
{\it Class. Quantum Grav.} \textbf{22} (2005) 2135.
\bibitem{JR05subm}   Jaekel M.-T. and Reynaud S.,
{\it submitted} (2005), in \eprint{gr-qc/0510068}.
\bibitem{Jaekel95} Jaekel M. T. and Reynaud S., \textit{Annalen der Physik}
\textbf{4} (1995) 68.
\bibitem{Einstein07} Einstein A., \textit{Jahrbuch der Radioaktivit\"at und 
Elektronik} \textbf{4} (1907) 411.
\bibitem{Einstein11} Einstein A., \textit{Annalen der Physik} \textbf{35} (1911) 898.
\bibitem{Adelberger03} Adelberger E. G., Heckel B. R. and Nelson A. E., \textit{%
Ann. Rev. Nucl. Part. Sci.} \textbf{53} (2003) 77.
\bibitem{Williams96} Williams J. G., Newhall X. X. and Dickey J. O., \textit{%
Phys. Rev.} \textbf{D53} (1996) 6730.
\bibitem{Hellings83} Hellings R. W., Adams P. J., Anderson J. D. \textit{et al}, 
\textit{Phys. Rev. Lett.} \textbf{51} (1983) 1609.
\bibitem{Anderson96} Anderson J. D., Gross M., Nordtvedt K. L. and Turyshev S. G., 
\textit{Astrophys. J.} \textbf{459} (1996) 365.
\bibitem{Einstein15} Einstein A., \textit{Sitzungsberichte der Preussischen 
Akademie der Wissenschaften zu Berlin} (1915) 844.
\bibitem{Hilbert} Hilbert D., \textit{Nachrichten von der Gesellshaft der 
Wissenshaften zu G\"ottingen} (1915) 395.
\bibitem{Einstein16} Einstein A., \textit{Annalen der Physik} \textbf{49} (1916) 769.
\bibitem{Eddington} Eddington A. S., \textit{The mathematical theory of relativity}
(Cambridge U. P., Cambridge, 1957).
\bibitem{Robertson} Robertson H. P. in \textit{Space age astronomy} (Academic Press, 
1962).
\bibitem{Schiff66} Ross D. H. and Schiff L. I., {\it Phys. Rev.} {\bf 141} (1966) 1215.
\bibitem{Nordtvedt68} Nordtvedt K. Jr., {\it Phys. Rev.} {\bf 169} (1968) 1014 \& 1017. 
\bibitem{WillNordtvedt72} Will C. M. and Nordtvedt K., 
{\it Astrophys. J.} {\bf 177} (1972) 757; Nordtvedt K. and Will C. M., \ibid 775.
\bibitem{Shapiro04}  Shapiro S. S., Davis J. L., Lebach D. E. and Gregory J. S.,
{\it Phys. Rev. Lett.} {\bf 92} (2004) 121101.
\bibitem{Bertotti03} Bertotti B., Iess L. and Tortora P., {\it Nature} {\bf 425}
(2003) 374.
\bibitem{Talmadge88} Talmadge C.  \textit{et al},
 \textit{Phys. Rev. Lett.} \textbf{61} (1988) 1159.
\bibitem{LLR02} Nordtvedt K. Jr., \eprint{gr-qc/0301024}.
\bibitem{Coy03} Coy J., Fischbach E., Hellings R., Talmadge C. and Standish E.M., private communication (2003). 
\bibitem{Hoyle} Hoyle C.D. \textit{et al}, \textit{Phys. Rev. Lett.} \textbf{%
86} (2001) 1418; \textit{Phys. Rev. }\textbf{D70} (2004) 042004.
\bibitem{Long} Long J. \textit{et al}, \textit{Nature} \textbf{421}(2003) 922.
\bibitem{Chiaverini} Chiaverini J. \textit{et al}, \textit{Phys. Rev. Lett.} 
\textbf{90} (2003) 151101.
\bibitem{Bordag01} Bordag M., Mohideen U. and Mostepanenko V.M., \textit{%
Phys. Rep.} \textbf{353} (2001) 1.
\bibitem{Lambrecht02} Lambrecht A. and Reynaud S., in \textit{ Poincar\'{e}
Seminar 2002}, ed. B. Duplantier and V. Rivasseau (Birkha\"{u}ser Verlag, Basel, 2003), p. 109.
\bibitem{Decca03} Decca R.S.  \etal, \textit{Phys. Rev. }\textbf{D68} (2003) 116003.
\bibitem{Chen04} Chen F., \textit{et al}, \textit{Phys. Rev. }\textbf{A69} (2004) 022117.
\bibitem{Reasenberg79} Reasenberg R.D. \textit{et al}, \textit{Astrophys. J.} 
\textbf{234} (1979) L219.
\bibitem{Kolosnitsyn03} Kolosnitsyn N.I. and Melnikov V.N., \textit{Gen. Rel. Grav.}
\textbf{36} (2004) 1619.
\bibitem{JR04}  Jaekel M.-T. and Reynaud S.,
{\it Int. J. Mod. Phys.} \textbf{A20} (2005) 2294.
\bibitem{Damour} Damour T., {\it Class. Quantum Grav.} {\bf 13} (1996) A33;
Damour T., Piazza F. and Veneziano G., {\it Phys. Rev.} {\bf D 66} (2002) 046007.
\bibitem{Overduin00} Overduin J. M., {\it Phys. Rev.} {\bf D 62} (2000) 102001.
\bibitem{Landau} Landau L. D. and Lifschitz E. M., \textit{The classical theory of 
fields} (Butterworth, 1975).
\bibitem{MTW} Misner C. W., Thorne K. S. and Wheeler J. A., 
Gravitation (Freeman, 1972).
\bibitem{Nieto05early} Nieto M. M. and Anderson J. D., \textit{Using early data to
illuminate the Pioneer anomaly} (2005) \eprint{gr-qc/0507052}.
\bibitem{TuryshevO5grex} Turyshev S. G., \textit{Recent progress in the study
of the Pioneer anomaly} (GREX meeting, 2005) 
{http://www.spectro.jussieu.fr/GREX/Paris05/Talks/Turyshev.pdf}.
\bibitem{Lisa} LISA Site @ ESA {http://www.rssd.esa.int/Lisa}.
\bibitem{Hipparcos} HIPPARCOS Site @ ESA {http://www.rssd.esa.int/Hipparcos}.
\bibitem{LATOR05} LATOR Collaboration: S.G. Turyshev, H. Dittus \etal
{\it Fundamental Physics with the Laser Astrometric Test of Relativity} {gr-qc/0506104}.
\bibitem{Gaia} GAIA Site @ ESA {http://www.rssd.esa.int/Gaia}.
\end{references}
\def\etal{\textit{et al }}
\def\ibid{\textit{ibidem }}

\end{document}